\documentclass[aps, prl, twocolumn, nofootinbib, floatfix, superscriptaddress]{revtex4-1}

\usepackage{graphicx}% Include figure files
\usepackage{dcolumn}% Align table columns on decimal point
\usepackage{bm}% bold math
\usepackage{subfigure}
\usepackage{epsfig}
\usepackage{amsmath}
\usepackage{amsfonts}
\usepackage{graphicx}
\usepackage{amssymb}
\usepackage{nccmath}
\usepackage{amssymb,amsmath,amsfonts}
\usepackage{xfrac}
\usepackage{color}
\usepackage{tikz}
\usepackage[T1]{fontenc}
\usepackage[utf8]{inputenc}
\usepackage{ulem}

\usepackage{tcolorbox}
\usepackage{hyperref}
\hypersetup{colorlinks=true, citecolor=red, linkcolor=blue, urlcolor=blue}

\definecolor{rossos}{cmyk}{0,1,1,0.55}
\definecolor{bluscuro}{rgb}{0.15, 0.2, .85}
\definecolor{bluchiaro}{cmyk}{1,.3,0.,0.1}

\newcommand{\OmegaPBH}{\Omega_{\mathrm{PBH}}}
\newcommand{\OmegaGW}{\Omega_{\mathrm{GW}}}

\let\oldsqrt\sqrt
\def\sqrt{\mathpalette\DHLhksqrt}
\def\DHLhksqrt#1#2{%
\setbox0=\hbox{$#1\oldsqrt{#2\,}$}\dimen0=\ht0
\advance\dimen0-0.2\ht0
\setbox2=\hbox{\vrule height\ht0 depth -\dimen0}%
{\box0\lower0.4pt\box2}}

\newcommand{\dd}{\mathrm{d}}
\newcommand{\sss}[1]{{\scriptscriptstyle{#1}}}
\newcommand{\boldmathsymbol}[1]{{\ensuremath{\boldsymbol{#1}}}}

\newcommand{\uPl}{\mathrm{Pl}}

\newcommand{\umin}{\mathrm{min}}
\newcommand{\umax}{\mathrm{max}}

\newcommand{\usssPl}{\sss{\uPl}}

\newcommand{\calH}{\mathcal{H}}

\newcommand{\Mp}{M_\usssPl}

\newcommand{\beq}{\begin{equation}}
\newcommand{\eeq}{\end{equation}}
\newcommand{\bea}{\begin{equation}\begin{aligned}}
\newcommand{\eea}{\end{aligned}\end{equation}}

\newlength{\wsingfig}
\newlength{\wdblefig}
\newlength{\wquadfig}
\newlength{\wtriplefig}
\newcommand{\Eq}[1]{Eq.~(\ref{#1})}

\newcommand{\Fig}[1]{Fig.~{\ref{#1}}}

\newcommand{\be}{\begin{equation}}

\begin{document}

 \title{Gravitational wave signatures of no-scale Supergravity in NANOGrav and beyond}

 \author{Spyros Basilakos}
% \email{svasil@Academyofathens.gr}
\affiliation{National Observatory of Athens, Lofos Nymfon, 11852 Athens, 
Greece}
\affiliation{Academy of Athens, Research Center for Astronomy and Applied 
Mathematics, Soranou Efesiou 4, 11527, Athens, Greece}
\affiliation{School of Sciences, European University Cyprus, Diogenes 
Street, Engomi, 1516 Nicosia, Cyprus}

\author{Dimitri~V.~Nanopoulos}
%\email{dimitri@physics.tamu.edu}
\affiliation{George P. and Cynthia W. Mitchell Institute for Fundamental
 Physics and Astronomy, Texas A\&M University, College Station, Texas 77843, 
USA}
 \affiliation{Astroparticle Physics Group, Houston Advanced Research Center 
(HARC),
  Mitchell Campus, Woodlands, Texas 77381, USA }
\affiliation{
  Academy of Athens, Division of Natural Sciences,
Athens 10679, Greece }

\author{Theodoros Papanikolaou}
\email{papaniko@noa.gr}
\affiliation{National Observatory of Athens, Lofos Nymfon, 11852 Athens, 
Greece}
\affiliation{Scuola Superiore Meridionale, Largo San Marcellino 10, I-80138, Naples, Italy}
\affiliation{Istituto Nazionale di Fisica Nucleare, Sezione di Napoli, Complesso Universitario di Monte S. Angelo, Via Cintia Edificio 6, 80126 Napoli, Italy}

\author{Emmanuel N. Saridakis}
%\email{msaridak@noa.gr}
 \affiliation{National Observatory of Athens, Lofos Nymfon, 11852 Athens, 
Greece}
\affiliation{CAS Key Laboratory for Researches in Galaxies and Cosmology, 
Department of Astronomy, University of Science and Technology of China, Hefei, 
Anhui 230026, P.R. China}

\author{Charalampos Tzerefos}
%\email{chtzeref@phys.uoa.gr}
 \affiliation{National Observatory of Athens, Lofos Nymfon, 11852 Athens, 
Greece}
 \affiliation{Department of Physics, National \& Kapodistrian University of 
Athens, Zografou Campus GR 157 73, Athens, Greece}

%\pacs{98.80.-k, 95.36.+x }

\begin{abstract}
In this Letter, we derive for the first time a characteristic three-peaked GW signal within the framework of no-scale Supergravity, being the low-energy limit of Superstring theory. %being the low-energy limit of Superstring theory.
We concentrate on the primordial gravitational wave (GW) spectrum induced due to second-order gravitational interactions by inflationary curvature perturbations as well as by isocurvature energy density perturbations of primordial black holes (PBHs) both amplified due to the presence of an early matter-dominated era (eMD) era before Big Bang Nucleosythesis (BBN). In particular, we work with inflection-point inflationary potentials naturally-realised within Wess-Zumino type no-scale Supergravity and giving rise to the formation of microscopic PBHs triggering an eMD era and evaporating before BBN. Remarkably, we obtain an abundant production of gravitational waves at the frequency ranges of $\mathrm{nHz}$, $\mathrm{Hz}$ and $\mathrm{kHz}$ and in strong agreement with Pulsar Time Array (PTA) GW data. Interestingly enough, a simultaneous detection of all three $\mathrm{nHz}$, $\mathrm{Hz}$ and $\mathrm{kHz}$ GW peaks %by PTA, ET/BBO and electromagnetic GW detectors respectively 
%\HT{ perhaps we can say " strengthens the plausibility of no-scale Supergravity models"} 
can constitute a potential observational signature for no-scale Supergravity.

\end{abstract}

\maketitle

%\section{Introduction}

%%%%%%%%%%%%%%%%%%%%%%%% Section 1:  Introduction %%%%%%%%%%%%%%%%%%%%%%%%%%%%%%%%%%%%%%%%%%%%%%%%%%%%%%%%%%%%%%%%
{\bf Introduction } -- According to the  15-year pulsar timing array data release of 
the NANOGrav Collaboration, there is positive evidence in favour of the 
presence of a low-frequency gravitational-wave (GW) background, which seems to be consistent with both cosmological~\cite{NANOGrav:2023hvm} and astrophysical~\cite{NANOGrav:2023hfp} interpretations. See here~\cite{Ellis:2023oxs} for a review for the different possible explanations of the PTA GW signal.

%Although this could be attributed to inspiraling supermassive black hole binaries \cite{NANOGrav:2023pdq,Ellis:2023dgf}, there seems to be strong indications that it might be more efficiently related to cosmological mechanisms of primordial origin, even more restricted than the possibilities left open by the previous 12.5-year data release \cite{NANOGrav:2023hvm}.

In this Letter, we 
provide a robust mechanism for the generation of such a signal within the framework of no-scale 
Supergravity 
\cite{Cremmer:1983bf,Ellis:1983sf,Ellis:1984bm,Lahanas:1986uc,Freedman:2012zz}, being the low-energy limit of Superstring theory~\cite{Witten:1985xb,Dine:1985rz,Antoniadis:1987zk}.
Such a construction provides in a natural way Starobinsky-like inflation \cite{Ellis:2013nxa,Kounnas:2014gda} with the 
desired observational features and with a consistent particle and cosmological phenomenology studied within the superstring-derived flipped SU(5) no-scale Supergravity \cite{Antoniadis:2020txn,Antoniadis:2021rfm}. Interestingly enough, through the aforementioned no-scale Supergravity construction, one obtains successfully as well for the first time ever the quark and charged lepton masses, which are actually  calculated directly from the string  \cite{Antoniadis:2020txn,Antoniadis:2021rfm}.

In the present study, we work within the framework of Wess-Zumino no-scale Supergravity \cite{Ellis:2013xoa} with naturally-realised inflection-point single-field inflationary potentials that can give rise to the formation of microscopic PBHs~\footnote{We mention that 
the current debate on PBH formation in single-field 
inflation models due to backreaction of small-scale one-loop corrections to the large-scale curvature power spectrum
\cite{Inomata:2022yte, Kristiano:2022maq,Choudhury:2023jlt,Choudhury:2023rks,Choudhury:2023vuj} has been evaded 
\cite{Franciolini:2023lgy,Firouzjahi:2023ahg} and PBH production in such 
scenarios is indeed viable.} with masses $M<10^9\mathrm{g}$, which can trigger early matter-dominated eras (eMD) before Big Bang Nucleosythesis (BBN) and address a plethora of cosmological issues among which the Hubble tension~\cite{Hooper:2019gtx,Papanikolaou:2023oxq}. Finally, we extract the stochastic gravitational-wave (GW) signals induced due to second-order gravitational interactions by inflationary adiabatic perturbations as well as by isocurvature induced adiabatic perturbations due to Poisson fluctuations in the number density of PBHs, which are resonantly amplified due to the presence of the aforementioned eMD era driven by them.

Notably, we find a three-peaked induced GW signal lying within the frequency ranges of $\mathrm{nHz}$, $\mathrm{Hz}$ and $\mathrm{kHz}$ and in strong agreement with the
recently released PTA GW data. The simultaneous detection of all three $\mathrm{nHz}$, $\mathrm{Hz}$ and $\mathrm{kHz}$ GW peaks by current and future GW detectors can constitute a potential observational signature for no-scale Supergravity.

%%%%%%%%%%%%%%%%%%%%%%%% Section 2:  No-scale Supergravity  %%%%%%%%%%%%%%%%%%%%%%%%%%%%%%%%%%%%%%%%%%%%%%%%%%%%%%%%%%%%%%%%
{\bf  No-scale Supergravity } --
In the most general $N=1$ supergravity theory   three
functions are involved: the K\"ahler potential $K$ (this is a Hermitian
function of the matter scalar field $\Phi^i$ and quantifies its geometry),
the superpotential
$W$ and the function $f_{ab}$, which are holomorphic functions of the fields. It is characterized by the action
\begin{equation}
\label{01}
\mathcal{S}= \int d^4 x \sqrt{-g} \left( K_{i \bar{j}} \partial_{\mu} \Phi^i 
\partial ^{\mu} \bar{\Phi} ^{\bar{j}}-V  \right),
\end{equation}
where we    set the reduced Planck mass $M_P= (8 \pi G)^{-1/2}=1$.
The  general form of the field metric is
\begin{equation}
K_{i \bar{j}}(\Phi,\bar{\Phi})= \frac{\partial^2 K}{\partial \Phi^ i \partial 
\bar{\Phi}^{\bar {j}} }  \,,
\label{2b}
\end{equation}
while the scalar potential reads as
\begin{equation}
V= e^{K}\left(	\mathcal{D}_{\bar {i}} \bar{W} K^{\bar{i} j} \mathcal{D} _j W - 
 3|W|^2\right)+\frac{\tilde g^2}{2}(K^i T^a \Phi_{i})^2\, ,
\label{3}
\end{equation}
where $i=\{\phi,T\}$, $K^{i \bar{j}} $ is the inverse K\"ahler metric and the 
covariant derivatives are defined as $
 \mathcal{D}_i W  \equiv  \partial_i W + K_i W $ and $
 \mathcal{D}^i W   \equiv  \partial^i W - K^i W $ (the last term in  (\ref{3}) 
is the $D$-term potential
and is set to zero  since the fields $\Phi_i$ are gauge singlets).
Moreover, we have defined    $K_i \equiv  \partial K/ \partial \Phi^i$ and
its complex conjugate $K^{\bar{i}}$. 
From~(\ref{01}) it is clear that the kinetic term $\mathcal{L}_{KE}=K_{i \bar{j}} 
\partial_{\mu} \Phi^i \partial ^{\mu} \bar{\Phi} ^{\bar{j}}$
needs to be  fixed.
 
We consider a no-scale supergravity model with two  chiral 
superfields $T$, $\varphi$,
that parametrize the noncompact $SU(2,1)/SU(2) \times U(1)$ coset space, 
with K\"ahler potential~\cite{Ellis:1984bm,Nanopoulos:2020nnh}
\begin{equation}
K= -3 \ln  \left[ T +\bar{T} -\frac{\varphi \bar{\varphi}}{3}+a \, 
e^{-b(\varphi+ \bar{\varphi})^2}(\varphi+\bar{\varphi} )^4 \right] \, ,
\label{a1}
\end{equation} 
where $a$ and $b$ are real constants.
Now, the simplest globally supersymmetric model is the Wess-Zumino one, which 
has a single chiral superfield $\varphi$, and it involves a mass term $\hat{\mu}$ 
and a trilinear coupling $\lambda$, while the corresponding superpotential is 
  \cite{Ellis:2013xoa}
 \begin{equation}
W= \frac{\hat \mu}{2} \varphi^2  - \frac{\lambda}{3} \varphi^3.
 \label{1a}
 \end{equation}
 In the limit  $a=0$, and by matching the  
  $T$ field   to  the modulus  field and the $\varphi$ to   the inflaton field, 
one can  derive  a class of  no-scale theories 
that yield  Starobinsky-like effective potentials  
\cite{Ellis:2013xoa,Ellis:2013nxa}, where the  potential is 
calculated  along the real inflationary direction defined by
\begin{align}
T= \bar{T}= \frac{c}{2} \,\, ,  \quad \mathrm{Im}\varphi=0  \, ,
\label{eq:realdir}
\end{align} with $\lambda / \mu = 1/3$ and  $\hat{\mu}=\mu \sqrt{c/3}$, with $c$  a constant~\footnote{We should note here that, as it was shown in~\cite{Ellis:2013xoa,Ellis:2013nxa,Ellis:1984bs}, the stabilization of the $T$-field is always possible while preserving the no-scale structure and reducing to a Starobinsky-like model in a suitable limit, hence justifying our choice within the current analysis to treat the $T$-field as non-dynamical.}. 
In particular,  transforming through 
$
\varphi= \sqrt{3\, c} \,  \tanh \left( \frac{\chi }{\sqrt{3}}  \right)$ one 
recovers the Starobinsky potential, namely $ V (\chi)=\frac{\mu^2}{4} \left(1- 
e^{-\sqrt{\frac{2}{3}}   \, { \chi} } \right)^2 $.

First, we verify  the stability along the 
inflationary direction and then we recast the kinetic term in canonical 
form. Furthermore,  defining  $\mathrm{Re}\,  \varphi \equiv \, \phi$, 
the relevant term in  the action 
is      $K _{\varphi \bar{\varphi}}$, which  along the direction 
(\ref{eq:realdir}) is equal to $K_{\phi \phi}$, thus leading to 
%\beq\label{eq:dchi_dphi}
$\frac{d \chi}{d \phi}= \sqrt{2 K_{\phi \phi }}$.
%\eeq
Integrating the above equation, we find
\begin{widetext}
  \bea
K_{\phi \phi}= \frac{9 \left\{768 a ^2 \phi ^6 \left(2 b \phi ^2+1\right)+16 a 
\phi ^2 e^{4 b \phi^2} \left\{c \left[6 b \phi ^2
   \left(9-8 b \phi ^2\right)-9\right]+2 b \phi^4 \left(8 b \phi 
^2-5\right)\right\}+c e^{8 b \phi^2}\right\}}
{\left[e^{4 b
   \phi ^2} \left(\phi ^2-3 c\right)-48 a \phi ^4\right]^2},
\eea
\beq\label{eq:V_potential}
V(\phi) = \frac{3 e ^ {12 b \phi^2} \phi^2(c \mu^2 -2\sqrt{3c} \lambda \,  \mu 
\,  \phi+3 \lambda^2 \,  \phi^2)}
     {\left[  -48 a \phi^4 +e ^{4b\phi^2}(-3c+\phi^2) \right]^2\, 
     \left\{ 
e^{4b\phi^2} -  24\, a \, \phi^2[6+4b \, \phi^2(-9+8 b \, \phi^2)] \right\}},
\eeq
 \end{widetext}
while the last expression has been extracted working   at the real 
inflationary direction where $\rm{Im}(\phi)=0$ and $T=\bar{T}=c/2$.

%%%%%%%%%%%%%%%%%%%%%%%% Section 3: Inflationary Dynamics %%%%%%%%%%%%%%%%%%%%%%%%%%%%%%%%%%%%%%%%%%%%%%%%%%%%%%%%%%%%%%%%
{\bf Inflationary Dynamics} --
%\label{sec:inflation}
Let us now recast the inflationary dynamics both at the 
background and the perturbative level.
 Working in a flat Friedmann-Lema\^itre-Robertson-Walker (FLRW) background, the 
background metric   reads as $\mathrm{d}s^2 = -\mathrm{d}t^2 
+a^2(t)\mathrm{d}x^i\mathrm{d}x_i$ and the Friedmann equations   have the 
usual form:
$
H^2  = \frac{1}{3 }\left[\frac{\dot{\chi}^2}{2} + 
V\left(\phi(\chi)\right)\right] $ and $
\dot{H}   = - \frac{\dot{\chi}^2}{2}$,
with the inflationary potential $V(\chi)$ given by \Eq{eq:V_potential}, while 
the non-canonical field $\phi$ is expressed in terms of the canonical 
inflaton field $\chi$ through $
\frac{d \chi}{d \phi}= \sqrt{2 K_{\phi \phi }}$, and as 
usual 
$
\ddot{\chi}+3H\dot{\chi} + V_\chi= 0$.

As numerical investigation shows,  the inflaton field is constant for a few 
e-folds, which is expected since  the
inflationary potential  presents an inflection point around the inflaton's 
plateau value where 
$\mathrm{d}V/\mathrm{d}\chi =\mathrm{d}^2V/\mathrm{d}\chi^2 \simeq 0$, thus leading 
to a transient ultra-slow-roll (USR) period. In particular, during this USR 
phase the non-constant mode of the curvature fluctuations, which in standard 
slow-roll inflation would 
decay, actually grows exponentially, hence enhancing the curvature power spectrum at 
small scales, collapsing to form PBHs. This is a pure result of the extended K\"ahler
potential introduced in (\ref{a1}). We also found that for a viable choice of the theoretical parameters at hand, the inflationary potential (\ref{eq:V_potential}) gives rise to a spectral index $n_\mathrm{s}\simeq 0.96$ and a tensor-to-scalar ratio $r<0.04$ in strong agreement with the Planck data~\cite{Planck:2018vyg}.

Focusing now at the perturbative level and working with the 
comoving curvature perturbation defined as
$
\mathcal{R} \equiv \Phi + \frac{H}{\dot{\chi}}\delta \chi
$ (with $\Phi$ being the Bardeen potential of scalar perturbations),
we derive the Mukhanov-Sasaki (MS) equation reading as~\cite{Mukhanov:1990me} 
\beq\label{eq:MS_R_k}
\mathcal{R}^{\prime\prime}_k + (3+\epsilon_2-\epsilon_1)\mathcal{R}^\prime_k + 
\frac{k^2}{a^2H^2}\mathcal{R}_k = 0,
\eeq
where $\prime$ denotes differentiation with respect to the e-fold number and 
$\epsilon_1$,$\epsilon_2$ stand for the usual Hubble flow slow-roll (SR) 
parameters, while the curvature power spectrum  is defined as
\beq\label{eq:P_R}
\mathcal{P}_{\mathcal{R}}(k) \equiv 
\left(\frac{k^{3}}{2\pi^{2}}\right)|\mathcal{R}_{_k}|^{2}.
\eeq

\begin{figure}[h!]
\begin{center}
\includegraphics[width=0.52\textwidth]{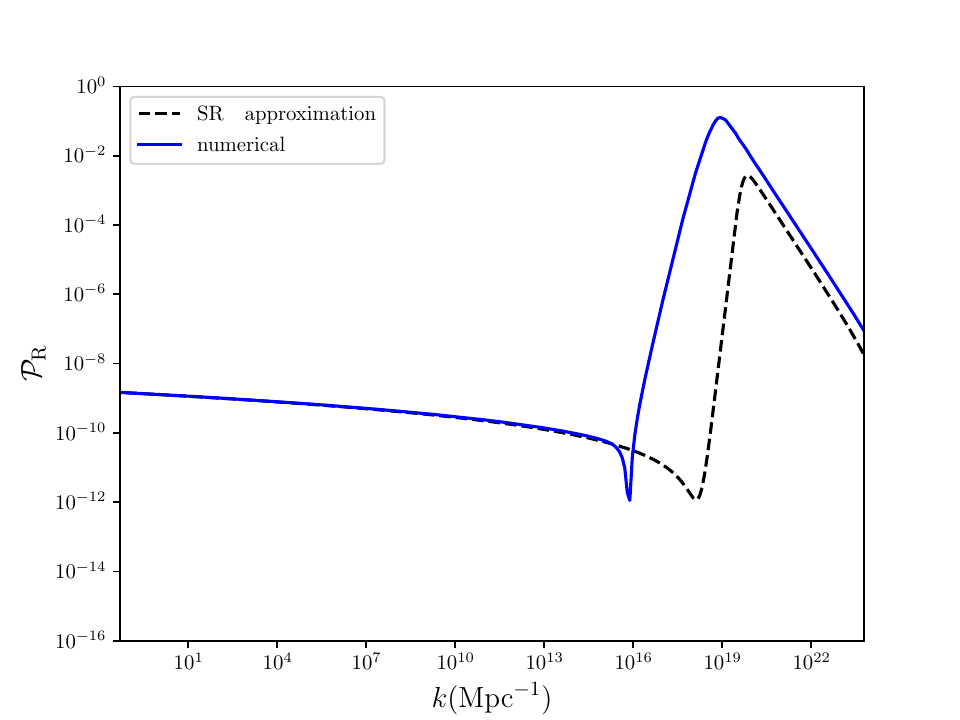}
\caption{{\it{The curvature power spectrum
$\mathcal{P}_\mathcal{R}(k)$  as a function of the wavenumber $k$, for 
$a=-1$, $b=22.35$, 
$c=0.065$, $\mu = 2\times 10^{-5}$, $\lambda/\mu = 0.3333449$ and $\phi_0 = 0.4295$ in Planck 
units. The black 
dashed curve represents the slow-roll (SR) approximation for 
$\mathcal{P}_\mathcal{R}(k)$, 
while the blue solid curve is the exact one after the numerical integration 
of the Mukhanov-Sasaki equation. }}}
\label{fig:P_RR}
\end{center}
\end{figure}

After numerical integration of \Eq{eq:MS_R_k} and using the 
Bunch-Davies vacuum initial conditions  on subhorizon 
scales, one can insert the solution of \Eq{eq:MS_R_k}  into (\ref{eq:P_R}) to 
obtain $\mathcal{P}_{\mathcal{R}}(k)$.
In \Fig{fig:P_RR} we present the obtained curvature power spectrum 
$\mathcal{P}_\mathcal{R}(k)$ for some fiducial values of the theoretical 
parameters involved, namely $a=-1$, $b=22.35$, 
$c=0.065$, $\mu = 2\times 10^{-5}$ and $\lambda/\mu = 0.3333449$ (we remind the reader 
that the value $\lambda/\mu = 1/3$ alongside $a=0$, corresponds 
to Starobinsky model). The initial value of the $\phi$ field was taken as $\phi_0 = 0.4295$ in Planck units. Very interestingly, as we can see from \Fig{fig:P_RR}, the curvature power 
spectrum   can 
be enhanced on small scales compared to the ones accessed by Cosmic 
Microwave Background (CMB) and Large-Scale Structure (LSS) probes, 
consequently leading to PBH formation~\cite{Nanopoulos:2020nnh}. However, in contrast 
to \cite{Nanopoulos:2020nnh}, in the 
current case we have $\lambda/\mu>1/3$ and thus we can produce   ultralight 
PBHs with masses less than $10^9\mathrm{g}$,  which evaporate before Big Bang 
Nucleosythesis (BBN). As one can observe from \Fig{fig:P_RR}, 
$\mathcal{P}_\mathcal{R}(k)$ peaks at $k_\mathrm{peak}\sim 
10^{19}\mathrm{Mpc}^{-1}$ which corresponds to a PBH mass forming in the
radiation-dominated era (RD) of the order of $
M_\mathrm{PBH}=17M_\odot\left(\frac{k}{10^6\mathrm{Mpc}^{-1}}\right)^{-2}\sim 
10^8\mathrm{g}$~\cite{Carr:2020gox} and evaporating at around $1\mathrm{MeV}$, i.e. BBN time.

At this point, we should comment as well on the fine-tuning of the ratio $\lambda/\mu$. As it was shown in~\cite{Ellis:2013nxa,Ellis:2018zya}, there is a unified and general treatment of Starobinsky-like inflationary  avatars of $SU(2,1)/SU(2)\times U(1)$ no-scale supergravity models. Further, it has been demonstrated that these different no-scale Supergravity models are equivalent and exhibit $6$ specific equivalence classes~\cite{Ellis:2018zya}. As such, it is not  inconceivable that the fine-tuning of $\lambda/\mu$ may be reduced or even be eliminated in other realizations of our proposed no-scale mechanism. The main point which should be stressed here is the fact that the CMB data favors Starobinsky-like models that are endemic in no-scale Supegravity theories, which emerge as generic low-energy effective field theories derived directly from the string [For a recent work on the topic see~\cite{Antoniadis:2020txn}]. Note also that PBH formation in single-field inflation demands in general fine-tuning of the inflationary parameters~\cite{Cole:2023wyx}.

%%%%%%%%%%%%%%%%%%%%%%%% Section 4:  Primordial black hole formation %%%%%%%%%%%%%%%%%%%%%%%%%%%%%%%%%%%%%%%%%%%%%%%%%%%%%%%%%%%%%%%%
{\bf Primordial black hole formation} --
We will recap briefly now the fundamentals of PBH formation. PBHs form  out of 
the collapse of local overdensity regions when   the energy density contrast of 
the collapsing overdensity becomes greater than a critical threshold 
$\delta_\mathrm{c}$~\cite{Harada:2013epa, Musco:2020jjb}. At the end, working 
within the context of peak theory~\cite{Bardeen:1985tr} one can 
straightforwardly show that the PBH mass function, defined as the energy density 
contribution of PBHs per logarithmic mass 
$\beta(M)\equiv\frac{1}{\rho_\mathrm{tot}} 
\frac{\mathrm{d}\rho_\mathrm{PBH}}{\mathrm{d}\ln M}$, is given 
by~\cite{Young:2019yug}
\beq\label{eq:beta_full_non_linear}
\beta(M) = 
\int_{\nu_\mathrm{c}}^{\frac{4}{3\sigma}}\mathrm{d}\nu\frac{\mathcal{K}}{4\pi^2}
\left(\nu\sigma - \frac{3}{8}\nu^2\sigma^2 - \delta_{\mathrm{c}}\right)^\gamma 
\frac{\mu^3\nu^3}{\sigma^3}e^{-\nu^2/2},
\eeq
with $\nu_\mathrm{c} = \delta_{\mathrm{c},l}/\sigma$ and
$\delta_{\mathrm{c},l}=\frac{4}{3}\left(1 -
\sqrt{\frac{2-3\delta_\mathrm{c}}{2}}\right)$ being the linear PBH formation threshold.  The parameters $\sigma^2$ 
and $\mu^2$ are the smoothed power spectrum and its first moment  defined as 
\begin{align}
\sigma^2 & \equiv 
\frac{4(1+w)^2}{(5+3w)^2}\int_0^\infty\frac{\mathrm{d}k}{k}(kR)^4 
\tilde{W}^2(k,R) \mathcal{P}_\mathcal{R}(k), \\
\mu^2 & \equiv 
\frac{4(1+w)^2}{(5+3w)^2}\int_0^\infty\frac{\mathrm{d}k}{k}(kR)^4 
\tilde{W}^2(k,R) \mathcal{P}_\mathcal{R}(k)\left(\frac{k}{aH}\right)^2,
\end{align}
with $w$ being the equation-of-state parameter of the dominant background component, and $\tilde{W}^2(k,R)$ the Fourier transform of the  Gaussian window function 
$
\tilde{W}(R,k) = e^{-k^2R^2/2}$~\cite{Young:2014ana,Young:2019osy}.

We should highlight here that in \Eq{eq:beta_full_non_linear} we have 
accounted for the non-linear relation between the energy density contrast 
$\delta$ and the comoving curvature perturbation $\mathcal{R}$ giving rise to 
an inherent primordial non-Gaussianity of the $\delta$ 
field~\cite{DeLuca:2019qsy,Young:2019yug}, as well as for the fact that the PBH 
mass is given by the  critical collapse scaling law 
$
M_\mathrm{PBH} = M_\mathrm{H}\mathcal{K}(\delta-\delta_\mathrm{c})^\gamma$
\cite{Niemeyer:1997mt, Musco:2008hv},
where $M_\mathrm{H}$ is the mass within the cosmological horizon at PBH 
formation time, and $\gamma$ is the critical exponent at the time of PBH 
formation (for PBH formation in the RD era  $\gamma\simeq 0.36$). 
Regarding the parameter $\mathcal{K}$ we work with its representative value 
$\mathcal{K}\simeq 4$~\cite{Musco:2008hv}, while concerning  the value of 
the PBH formation threshold $\delta_\mathrm{c}$, we accounted for its dependence on the shape 
of the collapsing curvature power spectrum. At the end, following the formalism developed in~\cite{Musco:2020jjb} we found it equal to $\delta_\mathrm{c}\simeq 0.505$.

\begin{figure*}[t!]
\begin{centering}
\includegraphics[width=0.495\textwidth]{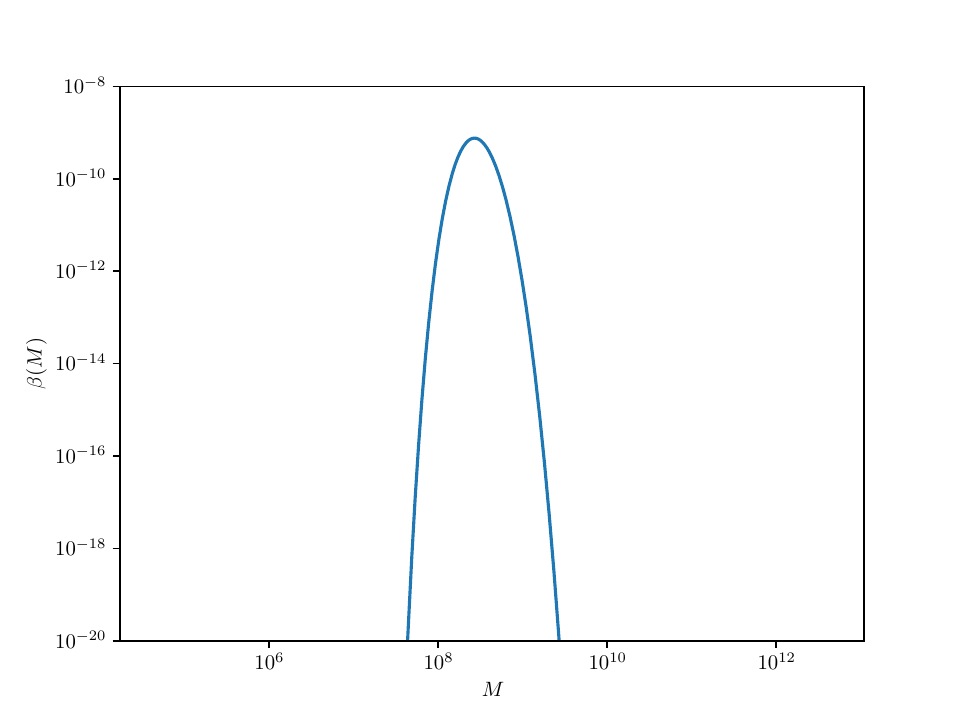}
  \includegraphics[width=0.495\textwidth, clip=true]
                  {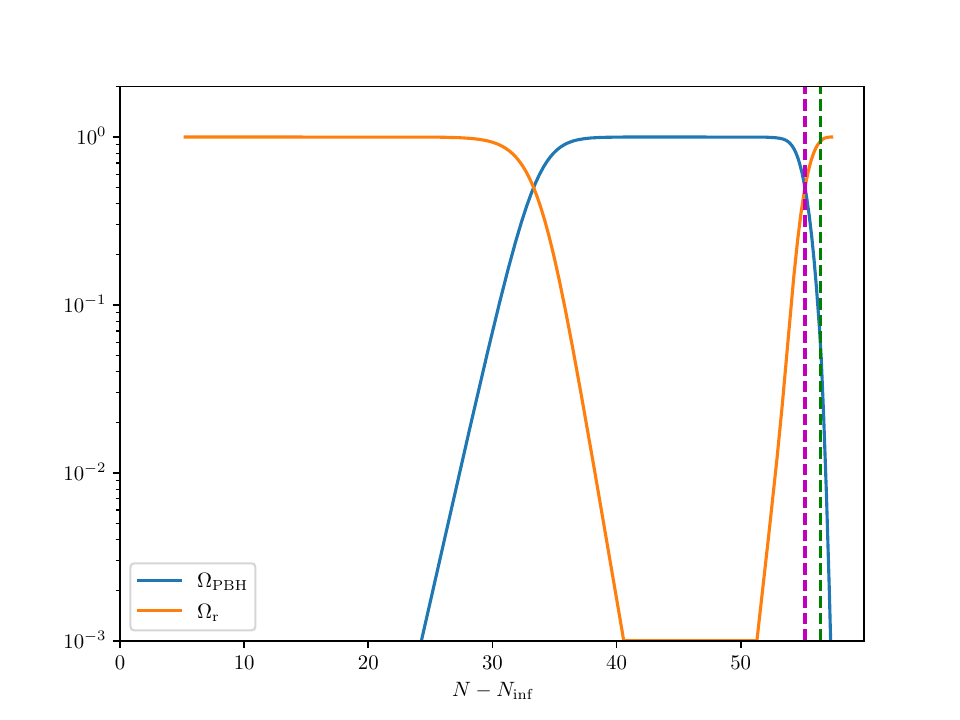}
  \caption{{\it{Left Panel: The PBH mass function at formation \Eq{eq:beta_full_non_linear}. Right Panel: The dynamical evolution of the background PBH and radiation
energy densities as a function of the e-folds passed from the end of inflation. The magenta 
vertical dashed line denotes 
the time of the onset of the radiation-dominated era, namely when 
$\Omega_\mathrm{r} = 0.5$, whereas the green dashed vertical line stands for 
the time when $\Omega_\mathrm{r} = 0.95$, namely when we are fully back to the 
radiation-dominated Universe. For both panels, we used $a=-1$, $b=22.35$, 
$c=0.065$, $\mu = 2\times 10^{-5}$, $\lambda/\mu = 0.3333449$ and $\phi_0 = 0.4295$ in Planck 
units. 
}}}
\label{fig:Omega_PBH}
\end{centering}
\end{figure*}

%%%%%%%%%%%%%%%%%%%%%% SECTION 5: %%%%%%%%%%%%%%%%%%%%%%%
{\bf The primordial black hole gas} -- 
Working within Wess-Zumino type 
no-scale supergravity \cite{Ellis:2013xoa}, we obtain an enhanced curvature power spectrum which is broader compared to the Dirac-monochromatic case (see \Fig{fig:P_RR}) but still sharp giving rise naturally to nearly monochromatic PBH 
mass distributions $\beta(M)$ [See the left panel of \Fig{fig:Omega_PBH}]. One then obtains in principle a ``gas'' of PBHs with different masses lying within the mass range 
$[10\mathrm{g},10^9\mathrm{g}]$, hence evaporating before 
BBN~\cite{Kawasaki:1999na}. Most of them however will have a common mass associated to the peak of the primordial curvature power spectrum. Due to the effect of 
Hawking radiation, each PBH will loose its mass with the dynamical evolution of 
the latter being given by 
$
M(t) = M_\mathrm{f}\left\lbrace
1-\frac{t-t_{\mathrm{f}}}{\Delta 
t_{\mathrm{evap}}(M_\mathrm{f})}\right\rbrace^{1/3}$
\cite{Hawking:1974rv},
where $t_{\mathrm{f}}$ is the PBH formation time and $\Delta t_{\mathrm{evap}}$ 
is the black hole evaporation time scaling with the black hole mass as
$
\Delta t_{\mathrm{evap}}(M_\mathrm{f}) =\frac{160}{\pi 
g_\mathrm{*}}\frac{M^3_\mathrm{f}}{\Mp^4}$,
with $g_\mathrm{*}$ being the effective number of relativistic degrees of 
freedom. 

If now $\bar{\beta}$ denotes
the mass fraction without accounting for Hawking evaporation, one can recast $\OmegaPBH (t)$ as
\begin{align}
\label{eq:OmegaPBH:continuous}
 \Omega_{\mathrm{PBH}}(t)  =
 \int_{M_\umin }^{M_\umax} \bar{\beta}
 \left(M,t\right)\left\lbrace
1-\frac{t-t_{\mathrm{ini}}}{\Delta 
t_{\mathrm{evap}}(M_\mathrm{f})}\right\rbrace^{1/3} \dd\ln M\, ,
\end{align}
where $t_{\mathrm{ini}}$ denotes the initial time in our dynamical evolution, 
which is basically the formation time of the smallest PBH mass considered. Regarding now 
the lower mass bound $M_\umin$, it will be given as the maximum between 
the minimum PBH mass at formation and the PBH mass evaporating at time $t$ 
defined as $M_\mathrm{evap}(t)\equiv \left( \frac{\pi g_\mathrm{*}\Mp^4 
\Delta 
t_\mathrm{evap}}{160}\right) ^{1/3}$. One then obtains that  $M_\umin= \max 
[M_\mathrm{f,min},M_\mathrm{evap}(t)]$.

After integrating numerically \Eq{eq:OmegaPBH:continuous} we 
obtain the PBH 
and the radiation background energy densities, which are depicted in the right panel of
\Fig{fig:Omega_PBH}. As we observe, the PBH abundance increases with 
time due to the effect of cosmic expansion, since at early times when 
$\OmegaPBH\ll 
1$ and Hawking radiation is negligible,  $\Omega_\mathrm{PBH}=
\frac{\rho_\mathrm{PBH}}{\rho_\mathrm{tot}}\propto a^{-3}/a^{-4}\propto a$, dominating in this way for a transient period the Universe's energy budget.
Then,
at some point Hawking evaporation becomes the driving force in the dynamics of 
$\OmegaPBH$ and the PBH abundance starts to decrease. Here, it is important to stress that, as one may notice from the right panel of \Fig{fig:Omega_PBH}, the transition from the eMD era driven by PBHs to the late RD (lRD) era lasts only one e-fold, hence one can treat the transition as instantaneous. This is related to the fact that the PBH mass function at formation can be treated as monochromatic. From the left panel of \Fig{fig:Omega_PBH} one can clearly see that the PBH mass function at formation $\beta(M)$ decays $11$ orders of magnitude in less than two decades in $M$.

We should mention at this point that the initial reheating temperature is determined by the energy scale at the end of inflation considering instantaneous reheating, i.e. $\rho_\mathrm{inf} = g_*\frac{\pi^2}{30}T^4_\mathrm{reh}$, where $ g_*$ is the effective number of relativistic degrees of freedom. For our inflationary setup, we find after solving numerically the Klein-Gordon equation, that inflation ends when $\rho^{1/4}_\mathrm{tot} = \rho_\mathrm{inf} = 10^{15}\mathrm{GeV}$. In contrast, the second reheating happens when PBHs evaporate, just before BBN, at around $1\mathrm{MeV}$.

%%%%%%%%%%%%%%%%%%%%%%%% Section 5:  Scalar induced gravitational waves %%%%%%%%%%%%%%%%%%%%%%%%%%%%%%%%%%%%%%%%%%%%%%%%%%%%%%%%%%%%%%%%

{\bf Scalar induced gravitational waves} -- 
Let us now derive the gravitational waves induced due to second order gravitational interactions by first order curvature perturbations~\cite{Matarrese:1992rp,Matarrese:1993zf,Matarrese:1997ay,Mollerach:2003nq} [See~\cite{Domenech:2021ztg} for a review]. Working in the Newtonian gauge~\footnote{We work within the Newtonian gauge as it is standardly used in the literature related to SIGWs.  The effect of the gauge dependence of the SIGWs is discussed in~\cite{Hwang:2017oxa,Tomikawa:2019tvi,DeLuca:2019ufz,Inomata:2019yww}.}  the 
perturbed metric is written as
$
\mathrm{d}s^2 = a^2(\eta)\left\lbrace-(1+2\Phi)\mathrm{d}\eta^2  + 
\left[(1-2\Phi)\delta_{ij} + 
\frac{h_{ij}}{2}\right]\mathrm{d}x^i\mathrm{d}x^j\right\rbrace$,
where $\Phi$ is the first order Bardeen gravitational potential and $h_{ij}$ stands for the 
second order tensor perturbation. Then, by 
performing a Fourier transform of the tensor perturbation, the equation of 
motion for $h_\boldmathsymbol{k}$ will be written as
\beq\label{Tensor eq}
h_\boldmathsymbol{k}^{s,\prime\prime} + 
2\mathcal{H}h_\boldmathsymbol{k}^{s,\prime} + k^{2} h^s_\boldmathsymbol{k} = 4 
S^s_\boldmathsymbol{k}, 
\eeq
where  $s = (+), (\times)$, $\mathcal{H}$ is the conformal Hubble parameter and while
the polarization tensors  $e^{s}_{ij}(k)$ are the standard  
ones~\cite{Espinosa:2018eve} and the source function 
$S^s_\boldmathsymbol{k}$ is given by~\footnote{We mention here that in this work we neglect possible effects of non-Gaussianities~\cite{Cai:2018dig} and one-loop corrections~\cite{Chen:2022dah} to the induced GW background.}
 \begin{eqnarray}
&&\!\!\!\!\!\!\!\!\!\!\!\!\!\!
S^s_\boldmathsymbol{k}  = \int\frac{\mathrm{d}^3 
\boldmathsymbol{q}}{(2\pi)^{3/2}}e^s_{ij}(\boldmathsymbol{k})q_iq_j\Big[
2\Phi_\boldmathsymbol{q}\Phi_\boldmathsymbol{k-q}  \nonumber\\
&& \!\!\!\!\!\!\!\!
+ 
\frac{4}{3(1\!+\!w_\mathrm{tot})}(\mathcal{H}^{-1}\Phi_\boldmathsymbol{q} 
^{\prime}+\Phi_\boldmathsymbol{q})(\mathcal{H}^{-1}\Phi_\boldmathsymbol{k-q} 
^{\prime}+\Phi_\boldmathsymbol{k-q}) \Big].
\label{eq:Source:def}
\end{eqnarray}
%The polarization tensors  $e^{s}_{ij}(k)$ are defined 
%as~\cite{Espinosa:2018eve}
%\beq
%e^{(+)}_{ij}(\boldmathsymbol{k}) = \frac{1}{\sqrt{2}}
%\begin{pmatrix}
%1 & 0 & 0\\
%0 & -1 & 0 \\ 
%0 & 0 & 0 
%\end{pmatrix}, \quad
%e^{(\times)}_{ij}(\boldmathsymbol{k}) = \frac{1}{\sqrt{2}}
%\begin{pmatrix}
%0 & 1 & 0\\
%1 & 0 & 0 \\ 
%0 & 0 & 0 
%\end{pmatrix}.
%\eeq
 
After a long but straightforward calculation, one obtains a tensor power 
spectrum  $\mathcal{P}_{h}(\eta,k)$ reading as~\cite{Ananda:2006af,Baumann:2007zm,Kohri:2018awv,Espinosa:2018eve}
 \begin{eqnarray}
&&
\!\!\!\!\!\!\!\!\!
\mathcal{P}^{(s)}_h(\eta,k) = 4\int_{0}^{\infty} 
\mathrm{d}v\int_{|1-v|}^{1+v}\mathrm{d}u\! \left[ \frac{4v^2 - 
(1\!+\!v^2\!-\!u^2)^2}{4uv}\right]^{2}\nonumber\\
&&
\ \ \ \ \ \ \ \ \ \ \ \ \ \ \ \ \ \ \ \ \ \ \ \ \ \ \cdot
I^2(u,v,x)\mathcal{P}_\Phi(kv)\mathcal{P}
_\Phi(ku)\,,
  \label{Tensor Power Spectrum}
\end{eqnarray}
where the two auxiliary variables $u$ and $v$ are defined as $u \equiv 
|\boldmathsymbol{k} - \boldmathsymbol{q}|/k$ and $v \equiv q/k$, and the kernel 
function $I(u,v,x)$ is a complicated function containing information for the 
transition between the eMD era driven by PBHs and the 
lRD era~\cite{Kohri:2018awv,Inomata:2019ivs,Inomata:2019zqy,Inomata:2020lmk,Papanikolaou:2022chm}. 
Hence, we can recast the GW spectral abundance 
defined as the GW energy density 
per logarithmic comoving scale as~\cite{Maggiore:1999vm,Kohri:2018awv}
\beq\label{Omega_GW}
\Omega_\mathrm{GW}(\eta,k)\equiv 
\frac{1}{\bar{\rho}_\mathrm{tot}}\frac{\mathrm{d}\rho_\mathrm{GW}(\eta,k)}{
\mathrm{d}\ln k} = 
\frac{1}{24}\left(\frac{k}{\calH(\eta)}\right)^{2}\overline{\mathcal{P}^{(s)}
_h(\eta,k)}.
\eeq
Finally,  considering that the 
radiation energy 
density reads as $\rho_r = 
\frac{\pi^2}{30}g_{*\mathrm{\rho}}T_\mathrm{r}^4$ and that the temperature of 
the primordial plasma $T_\mathrm{r}$ scales as $T_\mathrm{r}\propto 
g^{-1/3}_{*\mathrm{S}}a^{-1}$, one finds that the GW spectral abundance at 
our present epoch reads as
\beq\label{Omega_GW_RD_0}
\Omega_\mathrm{GW}(\eta_0,k) = 
\Omega^{(0)}_r\frac{g_{*\mathrm{\rho},\mathrm{*}}}{g_{*\mathrm{\rho},0}}
\left(\frac{g_{*\mathrm{S},\mathrm{0}}}{g_{*\mathrm{S},\mathrm{*}}}\right)^{4/3}
\OmegaGW(\eta_\mathrm{*},k),
\eeq
where $g_{*\mathrm{\rho}}$ and $g_{*\mathrm{S}}$ denote the energy and 
entropy relativistic degrees of freedom. Note that the reference 
conformal time $\eta_\mathrm{*}$ in the case of an instantaneous transition from the eMD to the lRD era should be of 
$\mathcal{O}(1)\eta_r$ \cite{Inomata:2019ivs,Inomata:2020lmk}. In the case of gradual transition,
$\eta_\mathrm{*}\sim (1-4)\eta_\mathrm{r}$ in order for $\Phi$ to have 
sufficiently decayed and the tensor modes to be considered as freely propagating GWs~\cite{Inomata:2019zqy,Papanikolaou:2022chm}.

%%%%%%%%%%%%%%%%%%%%%%%% Section 6:  Scalar induced gravitational waves %%%%%%%%%%%%%%%%%%%%%%%%%%%%%%%%%%%%%%%%%%%%%%%%%%%%%%%%%%%%%%%%

{\bf The relevant gravitational-wave sources} -- 
We can now concentrate on the different sources of GWs considered within this work. In particular, for a sharply peaked primordial curvature power spectrum like ours, GWs are sourced through two different mechanisms~\cite{Bhaumik:2022pil,Bhaumik:2022zdd}: Firstly, by the primordial inflationary curvature perturbations during the early RD (eRD) era, and later by early isocurvature PBH Poisson fluctuations during the eMD and the lRD eras. Given now the suddenness of the transition from the eMD to the lRD era, the induced GWs are resonantly amplified due to the large amplitude of the oscillations of the curvature perturbations after the sudden transition~\cite{Inomata:2019ivs,Inomata:2020lmk,Domenech:2020ssp}.

\begin{figure}[h!]
\begin{center}
\includegraphics[width=0.52\textwidth]{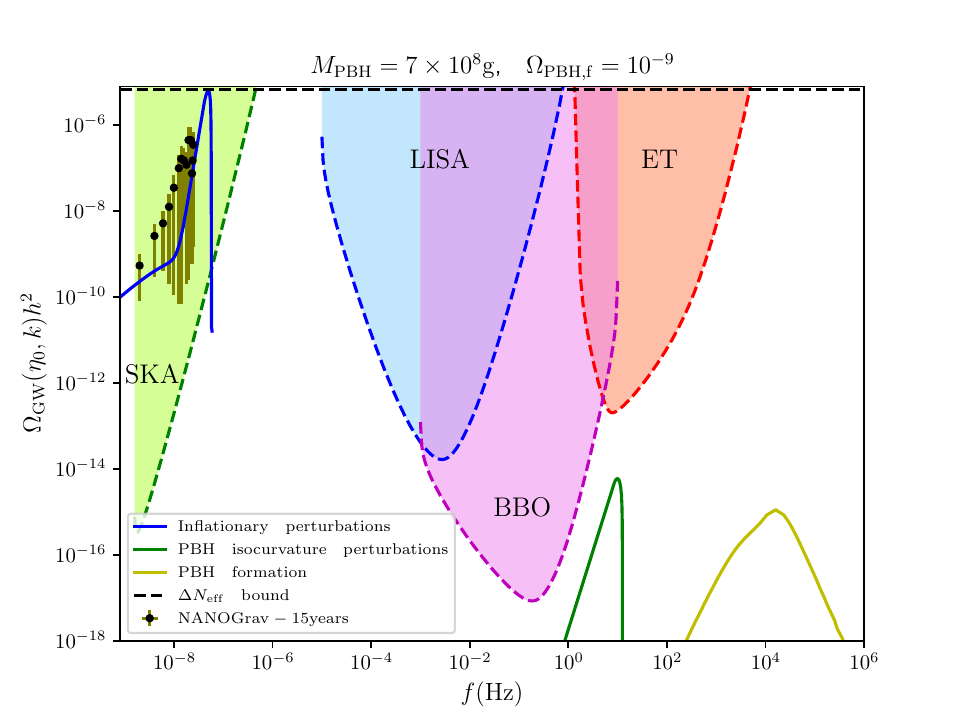}
\caption{ 
{\it{The stochastic three-peaked GW background induced by inflationary adiabatic (solid blue and yellow curves) and PBH isocurvature perturbations (solid green curve) arising from Wess-Zumino no-scale Supergravity with the extended K\"ahler potential (\ref{a1}) for $a=-1$, $b=22.35$, $c=0.065$, $\mu = 2\times 10^{-5}$, $\lambda/\mu = 0.3333449$ and $\phi_0 = 0.4295$ in Planck units and corresponding to $M_\mathrm{PBH}=7\times 10^8\mathrm{g}$ and $\Omega_\mathrm{PBH,f}= 10^{-9}$. On the top of our theoretical prediction for the induced stochastic GW background we show the  15-year NANOGrav GW data, as well as the sensitivities of SKA~\cite{Janssen:2014dka}, LISA~\cite{LISACosmologyWorkingGroup:2022jok}, BBO~\cite{Harry:2006fi} and ET~\cite{Maggiore:2019uih} GW experiments. In the horizontal black dashed line, we show also the upper bound on $\Omega_\mathrm{GW,0}h^2\leq 6.9\times 10^{-6}$ coming from the upper bound constraint on $\Delta N_\mathrm{eff}$ from CMB and BBN observations~\cite{Smith:2006nka}.} 
}}
\label{fig:GW_signals}
\end{center}
\end{figure}

To be more explicit, the first GW production mechanism gives rise to two GW peaks~\cite{Bhaumik:2023wmw}. The first peak is related to GWs induced by the enhanced primordial curvature power spectrum around the PBH scale, namely around $k=10^{19}\mathrm{Mpc}^{-1}$, which is associated to PBH formation. In order to extract this GW spectrum, one needs to use the kernel function $I(u,v,x)$ during an eRD era~\cite{Kohri:2018awv} when the PBHs form  and evolve the GW spectral abundance through the subsequent eMD era driven by PBHs, during which the GW spectrum is diluted as $\Omega_\mathrm{GW}\propto \rho_\mathrm{GW}/\rho_\mathrm{PBH}\propto a^{-4}/a^{-3}\propto 1/a$. At the end, one obtains that the induced GW due to PBH formation can be recast as
\beq\label{eq:Omega_GW_PBH_formation}
\Omega^{\mathrm{form}}_\mathrm{GW}(\eta_0,k) = \frac{a_\mathrm{d}}{a_\mathrm{evap}}c_g\Omega^{(0)}_\mathrm{r} \Omega_\mathrm{GW}(\eta_\mathrm{f},k),
\eeq
where $\eta_\mathrm{f}$ is the conformal PBH formation time, $c_g =\frac{g_{*\mathrm{\rho},\mathrm{*}}}{g_{*\mathrm{\rho},0}}
\left(\frac{g_{*\mathrm{S},\mathrm{0}}}{g_{*\mathrm{S},\mathrm{*}}}\right)^{4/3} \sim 0.4$ and $a_\mathrm{d}$ and $a_\mathrm{evap}$ are respectively the scale factors at the onset of eMD era when PBHs dominate and the lRD era when PBHs evaporate. $\Omega_\mathrm{GW}(\eta_\mathrm{f},k)$ is derived from \Eq{Omega_GW} at PBH formation time. One may naively expect from \Eq{Tensor Power Spectrum} and \Eq{Omega_GW} that for sharply peaked primordial curvature power spectra as in our case: $\Omega_\mathrm{GW}\propto \mathcal{P}^2_\Phi \propto \mathcal{P}^2_\mathcal{R}$, since $\mathcal{R}=2\Phi/3$ in superhorizon scales~\cite{Wands:2000dp}. At the end, for our fiducial choice of the inflationary parameters involved, the GW signal associated to PBH formation peaks at the $\mathrm{kHz}$ frequency range [See the yellow solid curve in \Fig{fig:GW_signals}]. %and is potentially detectable by electromagnetic GW detectors~\cite{Domcke:2023qle}.

Regarding now the second peak at $\mathrm{nHz}$, it is related to the resonant amplification of the curvature perturbation on scales entering the cosmological horizon during the eMD. In particular, the source of the enhancement leading to the $\mathrm{nHz}$ peak is the sudden transition from the eMD era to the lRD era. Specifically, during the transition the time derivative of the Bardeen potential goes very quickly from $\Phi^\prime = 0 $ (since in a MD era $\Phi = \mathrm{constant}$ ) to $\Phi^\prime \neq 0$ in the late RD era [See~\cite{Inomata:2019ivs,Domenech:2021ztg} for more details.]. This entails a resonantly enhanced production of GWs sourced mainly by the $\mathcal{H}^{-2}\Phi^{\prime 2}$ term in \Eq{eq:Source:def}.
    
Furthermore, since the sub-horizon energy density perturbations during a MD era scale linearly with the scale factor, i.e. $\delta\propto a$, one should ensure working within the perturbative regime. For this reason, we set a non-linear scale $k_\mathrm{NL}$ by requiring that $\delta_{\mathrm{k_\mathrm{NL}}}(\eta_\mathrm{r}) = 1$. In particular, following the analysis of~\cite{Assadullahi:2009nf,Inomata:2020lmk}, one can show that the non-linear cut-off scale~\footnote{It is important to stress here that this non-linear cut-off points out actually the limit of our ability to perform perturbative calculations. If one wants to go beyond the perturbative regime, they need
to perform high-cost numerical simulations, which goes beyond the scope of this
work.} at which $\delta_{k_\mathrm{NL}}(\eta_r) = 1$ can be recast as
\begin{equation}
k_\mathrm{NL} \simeq \sqrt{\frac{5}{2}}\mathcal{P}^{-1/4}_\mathcal{R}(k_\mathrm{NL})\mathcal{H}(\eta_r).
\end{equation}

Since within no-scale Supergravity we predict a Starobinsky-like inflationary setup with $n_\mathrm{s} = 0.965$, we can assume as a first approximation a scale-invariant curvature power spectrum of amplitude $2.1\times 10^{-9}$ as imposed by Planck~\cite{Planck:2018vyg}, giving rise to $ k_\mathrm{NL} \simeq 470/\eta_\mathrm{r} \simeq 235 k_\mathrm{r}$~\cite{Inomata:2019ivs},where $k_\mathrm{r}$ is the comoving scale crossing the cosmological horizon at the onset of the lRD era. Ultimately, the peak frequency of this $\mathrm{nHz}$ signal is associated with the non-linear comoving cut-off scale, $k_\mathrm{NL}=235 k_{r}$, which depends on the PBH mass $M$ as~\cite{Bhaumik:2023wmw} $k_\mathrm{r} = 2.1 \times 10^{11}\left(\frac{M}{10^4\mathrm{g}}\right)^{-3/2}\mathrm{Mpc}^{-1}$~\cite{Bhaumik:2023wmw}. The peak frequency of this signal can now be calculated from the formula $f_\mathrm{GW} = \frac{c  k_{\mathrm{NL}}}{2\pi a_0}$, with $c$ the speed of light and $a_0=1$. The end result is a $f$ peaking at $\mathrm{nHz}$.

\begin{figure}[h!]
\hspace{-0.8cm}
\includegraphics[width=0.52\textwidth]{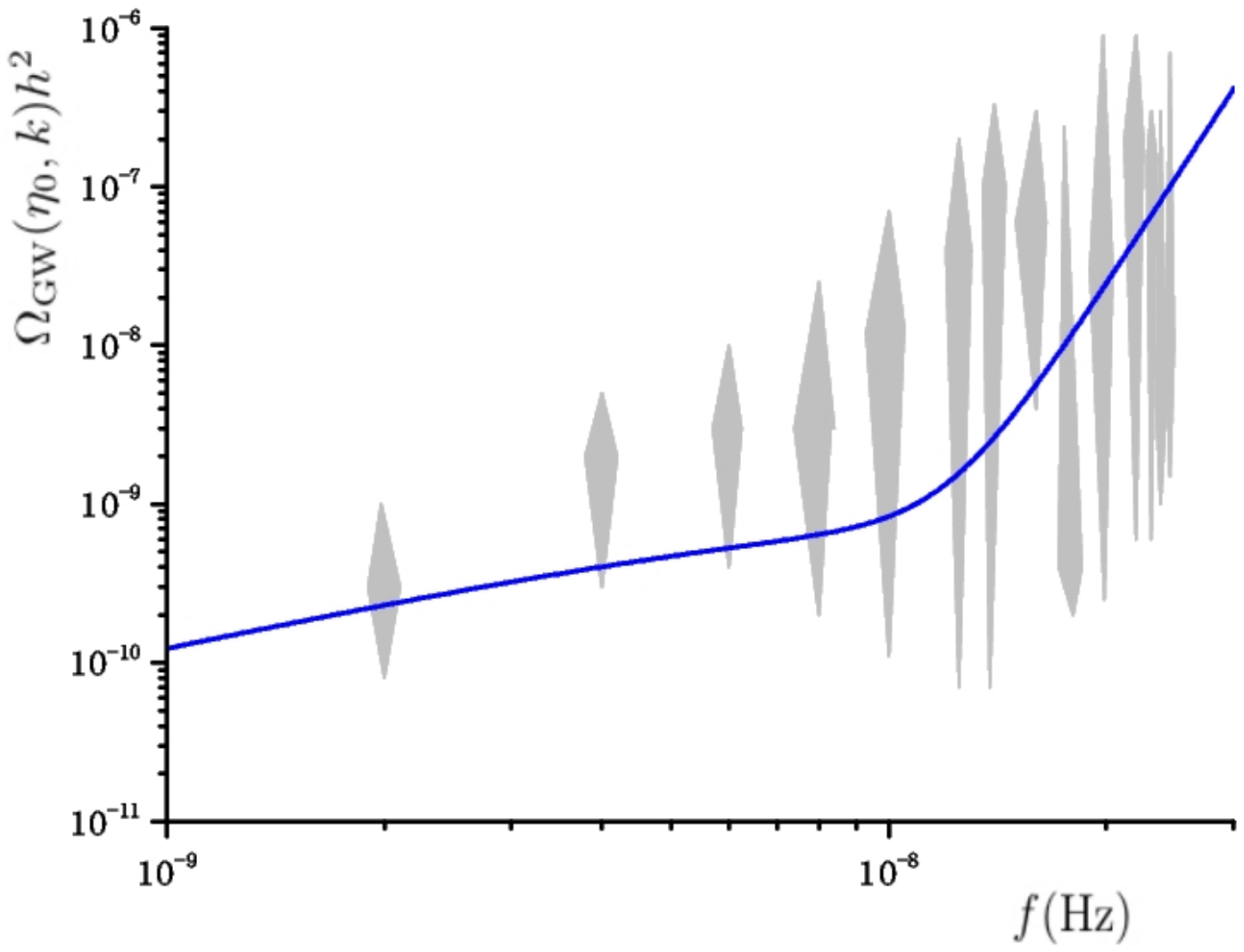}
\caption{ {\it{The stochastic GW background (GW density 
parameter as a function of the frequency) induced by resonantly amplified inflationary perturbations and 
arising from Wess-Zumino no-scale Supergravity with the extended K\"ahler potential (\ref{a1}) for  $a=-1$, $b=22.35$, 
$c=0.065$, $\mu = 2\times 10^{-5}$, $\lambda/\mu = 0.3333449$ and $\phi_0 = 0.4295$ in Planck units - corresponding to $M_\mathrm{PBH}=7\times 10^8\mathrm{g}$ and $\Omega_\mathrm{PBH,f}= 10^{-9}$ - on top of the 15-year NANOGrav data.
}}}
\label{fig:NANOGrav_comparison_figure}
\end{figure}

Remarkably, this second peak that corresponds to scales much larger than the PBH scale peaks at the $\mathrm{nHz}$ frequency range and is in strong agreement with the NANOGrav/PTA data - See the blue solid curve in \Fig{fig:GW_signals} as well as \Fig{fig:NANOGrav_comparison_figure} where we have zoomed in the NANOGrav frequency range. This second resonant peak is derived by 
\beq\label{eq:Omega_GW_resonance}
\Omega^{\mathrm{res}}_\mathrm{GW}(\eta_0,k) = c_g\Omega^{(0)}_\mathrm{r} \Omega_\mathrm{GW}(\eta_\mathrm{lRD},k),
\eeq
where $\eta_\mathrm{lRD}$ stands for a time during lRD by which the curvature perturbations decouple from the tensor perturbations and one is met with freely propagating GWs. $\Omega_\mathrm{GW}(\eta_\mathrm{lRD},k)$ is derived from \Eq{Omega_GW} where $I(u,v,x)=I_\mathrm{lRD}(u,v,x)$~\cite{Inomata:2019ivs}.

It is important to highlight here the possibility of later PBH formation due to the linear growth of the matter energy density perturbations during the PBH-driven eMD era. Interestingly enough, one expects a further PBH production similarly to PBH production from preheating~\cite{Martin:2020fgl} as well as early inflaton structure formation~\cite{Jedamzik:2010dq,Hidalgo:2022yed}. The study of this interesting phenomenology is beyond the scope of this work and will be studied elsewhere.

Finally, we consider the GW spectrum induced by the gravitational potential of our PBH population itself. To elaborate, %if one regards the PBH as a pressureless fluid, the process of PBH formation can be viewed as the transition of a fraction of the radiation background energy density into PBHs. 
assuming that PBHs are randomly distributed at formation time (i.e. they have Poisson statistics)~\cite{Desjacques:2018wuu,MoradinezhadDizgah:2019wjf}, their energy density is inhomogeneous while the total background radiation energy density is homogeneous. Therefore, 
the PBH energy density perturbation can be described by an isocurvature Poisson fluctuation~\cite{Papanikolaou:2020qtd} which in the subsequent PBH domination era will be converted into an adiabatic curvature perturbation associated to a PBH gravitational potential $\Phi_\mathrm{PBH}$. This gravitational potential $\Phi_\mathrm{PBH}$ will be another source of induced GWs.~\footnote{It is helpful to stress here that once we have a population of PBHs, it is the gravitational interaction between them (scattering, merging, etc.) that will entail the production of GWs. Within this work, we are interested in the large-scale counterpart of this signal, i.e. at distances much larger than the mean separation length between PBHs, for which PBHs can be viewed as a dust fluid with zero pressure. This fluid is characterised by its density perturbations, which can be treated within the framework of cosmological perturbation theory and which can induce GWs due to second order gravitational effects. This sets a UV cut-off scale $k_\mathrm{UV}$ which is actually  the mean PBH separation scale, below which one finds that $\mathcal{P}_{\delta_\mathrm{PBH}}(k)>1$, entering the non-linear regime. For more details see~\cite{Papanikolaou:2020qtd,Domenech:2020ssp}.}
The power spectrum of $\Phi_\mathrm{PBH}$ can be recast as~\cite{Papanikolaou:2020qtd}:
\beq
\label{eq:PowerSpectrum:Phi:PBHdom}
\mathcal{P}_{\Phi_\mathrm{PBH}}(k) = \frac{2}{3\pi} \left( \frac{k}{k_{\rm{UV}}} \right)^3 
\left(5+\frac{4}{9}\frac{k^2}{k_{\rm{d}}^2} \right)^{-2},
\eeq
where $k_{\mathrm{d}}$ stands for the comoving scale re-entering the 
cosmological horizon at PBH domination time and $k_\mathrm{UV}$ is a UV cutoff scale defined 
as $k_\mathrm{UV}\equiv a/\bar{r}$, where $\bar{r}$ corresponds to the mean PBH 
separation distance. One then can compute the relevant tensor power spectrum through \Eq{Tensor Power Spectrum} where now $\mathcal{P}_{\Phi}(k)$ should be replaced with $\mathcal{P}_{\Phi_\mathrm{PBH}}(k)$. One would expect that this GW signal will peak at $k_\mathrm{d}$,  namely the scale crossing the cosmological horizon at the onset of the PBH domination era, since as one may see from \Eq{eq:PowerSpectrum:Phi:PBHdom}, $\mathcal{P}_{\Phi_\mathrm{PBH}}(k)$ peaks at $k_\mathrm{d}$. However, due to a $k^8$ dependence of the kernel $I^2_\mathrm{lRD,res}(u,v,x_\mathrm{evap})$, the peak of the GW signal induced by PBH Poisson fluctuations is shifted from $k_\mathrm{d}$ to $k_\mathrm{UV}$. As one can see from the green solid line of \Fig{fig:GW_signals}, for our fiducial choice for the inflationary parameters at hand, this GW signal lies within the $\mathrm{Hz}$ frequency range with an amplitude of the order of $10^{-14}$, being close to sensitivity bands of the Einstein Telescope (ET)~\cite{Maggiore:2019uih} and Big Bang Observer (BBO)~\cite{Harry:2006fi}.

It is noteworthy that since GWs generated before BBN can act as an extra relativistic component, they will contribute to the effective number of extra neutrino species $\Delta N_\mathrm{eff}$, which is severely constrained by BBN and CMB  observations as $\Delta N_\mathrm{eff}< 0.3$~\cite{Planck:2018vyg}. This upper bound constraint on $\Delta N_\mathrm{eff}$ is translated to an upper bound on the GW amplitude which reads as~\cite{Smith:2006nka,Caprini:2018mtu} as $\Omega_\mathrm{GW,0}h^2\leq 6.9\times 10^{-6}$~\footnote{See~\cite{Domenech:2021wkk} for the effect on $\Delta N_\mathrm{eff}$ from PBH evaporation.}. This upper bound on $\Omega_\mathrm{GW}$ is shown with the horizontal black dashed line in \Fig{fig:GW_signals}.

 %%%%%%%%%%%%%%%%%%%%%%%% Section 7: Conclusions %%%%%%%%%%%%%%%%%%%%%%%%%%%%%%%%%%%%%%%%%%%%%%%%%%%%%%%%%%%%%%%%

{\bf Conclusions} -- 
In this Letter, we showed that no-scale Supergravity, being the low-energy limit of Superstring theory, seems to accomplish three main
achievements. Firstly, it provides a successful Starobinsky-like inflation realization with all the desired observational predictions regarding $n_\mathrm{s}$ and $r$~\cite{Antoniadis:2020txn,Ellis:2013xoa}. Secondly, it can naturally lead to inflection-point inflationary potentials giving rise to sharp mass distributions of microscopic PBHs triggering an eMD era before BBN, and thirdly it can induce through second order gravitational interactions a distinctive three-peaked GW signal.

In particular, working within the context of Wess-Zumino no-scale Supergravity we found i) a $\mathrm{nHz}$ GW signal induced by enhanced inflationary adiabatic perturbations and resonantly amplified due to the sudden transition from a eMD era driven by ``no-scale" microscopic PBHs to the standard RD era, being as well within the error-bars of the recently PTA GW data, ii) a $\mathrm{Hz}$ GW signal induced by the PBH isocurvature energy density perturbations and close to the GW sensitivity bands of ET and BBO GW experiments and iii) a $\mathrm{kHz}$ GW signal associated to the PBH formation. Remarkably, a simultaneous detection of all three $\mathrm{nHz}$, $\mathrm{Hz}$ and $\mathrm{kHz}$ GW peaks can constitute a potential observational signature for no-scale Supergravity.  

It is important to highlight here that we extracted the aforementioned three-peaked induced GW signal within the context of Wess-Zumino no-scale Supergravity. However, due to the unified treatment of Starobinsky-like inflationary avatars
of $SU(2,1)/SU(2)\times U(1)$ no-scale supergravity models~\cite{Ellis:2013nxa} which exhibit $6$ specific equivalence classes~\cite{Ellis:2018zya}, one expects that the three-peaked GW signal induced by inflationary adiabatic and PBH isocurvture perturbations will be a generic feature of any $SU(2,1)/SU(2)\times U(1)$ no-scale Supergravity theory with an appropriately deformed K\"ahler potential, e.g. of the form \Eq{a1}, leading to inflection-point inflationary potentials.

One should mention as well that in principle a three-peaked GW signal is a generic feature of any model predicting formation of ultra-light PBHs dominating the early Universe. However, the particular three-peaked GW signal found here, at $\mathrm{nHz}$, $\mathrm{Hz}$ and $\mathrm{kHz}$, is a  a pure byproduct of the no-scale Wess-Zumino inflection-point inflationary potential and consistent with the recently released PTA GW data. Thus, its detection will constitute a potential observational signature of no-scale Supergravity.

Finally, we need to note that in the case of broader PBH mass functions produced within no-scale Supergravity, one expects an oscillatory GW signal due to the gradualness of the transition from the eMD to the lRD era~\cite{Inomata:2019zqy,Papanikolaou:2022chm}. Such oscillatory GW signals have been proposed as well in other theoretical constructions [See~\cite{Braglia:2020taf,Fumagalli:2020nvq,Fumagalli:2021cel,Fumagalli:2021mpc,Mavromatos:2022yql}] and it will be quite tentative to distinguish between them experimentally. Finally, we should mention that a proper statistical comparison between no-scale Supergravity models and GW data will place strong constraints on the relevant parameter space involved. Such an analysis is in progress and will be published elsewhere.

{\bf Acknowledgements} -- 
The authors thank Ioanna Stamou, Valerie Domcke, Ninlanjandev Bhaumik,  Rajeev Kumar Jain and Marek Lewicki for useful and stimulating discussions. The work of DVN was supported
in part by the DOE grant DE-FG02-13ER42020 at Texas A\&M University and in part by the Alexander S. Onassis Public Benefit Foundation.
SB, ENS, TP and CT acknowledge the 
contribution of the LISA CosWG and the COST Actions  CA18108 ``Quantum Gravity Phenomenology in the multi-messenger approach''  and 
CA21136 ``Addressing observational tensions in cosmology with systematics and 
fundamental physics (CosmoVerse)''. TP and CT acknowledge as well financial support from the Foundation for Education and European Culture in Greece and A.G. Leventis Foundation respectively.

\bibliography{ref}

\end{document}